# Moiré based strain analysis in wurtzite GaAs – rock-salt (Pb,Sn)Te core/shell nanowires grown by molecular beam epitaxy

**Maciej Wojcik[1,3], Sania Dad[1], Piotr Dziawa[1,4], Slawomir Kret[1], Wojciech Pacuski[2], Janusz Sadowski[1,3]**

[1]*Institute of Physics, Polish Academy of Sciences, Aleja Lotnikow 32/46, PL-02-668 Warsaw, Poland*

[2]*Faculty of Physics, University of Warsaw, Pasteura 5, 02-093 Warsaw, Poland*

[3]*Ensemble3 Centre of Excellence, Wolczynska 133, 01-919 Warsaw, Poland*

[4]*International Research Center MagTop, Institute of Physics, Polish Academy of Sciences, Aleja Lotnikow 32/46, PL-02-668 Warsaw, Poland*

**Abstract**

We investigate core/shell GaAs/(Pb,Sn)Te nanowire nano-heterostructures with wurtzite (wz) GaAs cores and (Pb,Sn)Te topological crystalline insulator shells. The nanostructures have been grown by molecular beam epitaxy using two distinct MBE systems dedicated to III-V, and IV-VI semiconductors. The interface structure of wz-GaAs/(Pb,Sn)Te nanowires is investigated using high resolution transmission electron microscopy, scanning transmission electron microscopy and geometric phase analysis. Misfit dislocations and moiré fringes are observed as a direct result of the lattice mismatch between the core and the shell materials, and used to estimate strain in crystalline topological insulator shells. Our results point to a possibility of using moiré patterns analysis as an alternative, for estimating strain in the core-shell nanowire structures.

**1. Introduction**

Topological crystalline insulators (TCIs) belong to an extensively investigated research area of topological quantum matter. Interest in TCIs stems from the presence of topologically protected Dirac states on the high-symmetry surfaces of TCI crystals. The $Pb_{1-x}Sn_xTe$ substitutional solid solution is a representative example of TCIs, with band inversion occurring above some critical Sn content [1-6].

Although studies of the SnTe band structure date back to 1969, when an even number of band inversions were identified in the electronic structure at the four equivalent L points of the Brillouin zone, the material was initially classified as a trivial insulator. This was because an odd number of inversions is required for a material to be classified as a $Z_2$ strong topological insulator (TI) with time-reversal symmetry (TRS). However, this perspective shifted in 2012, when Hsieh et. al., proposed that these inversions of energy bands signify a mirror symmetry protected topological crystalline insulator (TCI) phase. They predicted that the coexistence of an inverted band structure and mirror symmetry leads to protected Dirac surface states. Dirac-like surface bands were observed on the (001) face of SnTe, and the presence of a metallic surface state, absent in the trivial counterpart PbTe, was evidenced[7]. Experimental validation swiftly followed in late 2012. Spin-resolved photoemission spectroscopy of $Pb_{1-x}Sn_xTe$ revealed four spin-polarised Dirac cones, consistent with the theoretical results predicting mirror-symmetry-protected TCI phase in SnTe and its Pb-based solid solutions[8]. At the same time, other groups observed TCI behaviour in $Pb_{1-x}Sn_xSe$. Adjusting the Sn content to $x \approx 0.23$, a temperature-driven topological phase transition from trivial to inverted bands was observed[9]. Together, these findings established the SnTe class as one of the first realisations of TCIs and paved the way for the discovery of higher-order topological insulators (HOTIs) and their associated lower-dimensional boundary modes[10].

The relatively large lattice parameters of rock-salt narrow bandgap IV-VI semiconductors, (in the range of 6.10 Å – 6.46 Å) prohibit use of typical substrates such as Si or GaAs for deposition of high quality epitaxial layers. However, unlike planar heterostructures, which are prone to high dislocation densities and microcracking, the high aspect ratio of NWs enables effective strain relaxation without high concentration of structural defects at interfaces [11] in the case of substantial lattice mismatch between heterostructure components. In wurtzite GaAs/(Pb,Sn)Te core/shell structures, the mismatch between the (0001) wz GaAs planes and the (001) rock-salt $Pb_{1-x}Sn_xTe$ planes is relatively modest (1.7% to 4%) along the growth axis[12]. Although the lateral mismatch is more significant, the shape anisotropy of the NW minimises misfit dislocations, resulting in structural quality that surpasses that of traditional thin-film growth methods[13-15].

The class of SnTe NWs provides a flexible material platform for exploring the interplay of topology and symmetry breaking[16-18]. The occurrence of low temperature (below 100 K) lower symmetry rhombohedral crystalline phase in SnTe with low concentration of carriers[19] enables investigations of HOTI states, i.e. 1D hinge states localized along crystalline edges[20], 0D corner states that emerge at the wire ends in the band-inverted insulator state, and potential Majorana zero modes[21]. In IV-VI NWs with square cross-sections, four sidewall surfaces and hinges can host these states[22, 23]. However, in GaAs/(IV-VI) core/shell nano-heterostructures with hexagonal cross-sections, this capacity extends to six[20, 24].

Core/shell nano-heterostructures offer significant functional advantages over pristine ones[25-30]. These include, strain engineering i.e. the core's band gap can be precisely tuned via lattice mismatch with the shell[31]. Furthermore it can enhance optoelectronic properties as IV-VI shells can suppress Auger recombination and surface trap states, thereby optimising charge transport and optical performance[25]. Features such as carrier multiplication make the IV-VI nanostructures suitable for use in LEDs, photodetectors and photovoltaics[32-34].

In this work we have studied III-V/IV-VI core/shell nano-heterostructures enabling formation of long nanowire-like structures comprising IV-VI TCI, which is challenging in the case of direct growth of $Pb_{1-x}Sn_xTe$ NWs using the MBE technique[35, 36]. We have used moiré patterns to estimate strain of the TCI material, which can influence the topological properties[37]. potentially leading to further improvement of the surface related properties.

## 2. Experimental

Core-shell NW nano-heterostructures have been grown by molecular beam epitaxy using two different MBE systems dedicated to III-V and IV-VI semiconductors, respectively. GaAs NWs are grown in the III-V MBE system on GaAs (111)B substrates with pre-deposited thin gold layer (5 Å Au deposited in a high vacuum sputtering system), in conditions favouring wz crystal structure. Then the substrates with GaAs NWs were transferred (through air) to the IV-VI MBE system and used for $Pb_{1-x}Sn_xTe$ shell deposition. Prior to the deposition of the shells, GaAs NWs were annealed to ~ 590 °C for thermal remove of native oxides from the sidewalls. This growth process is described in more details in Ref. 12. The shell thickness was chosen to be relatively low (about 10 nm) in order to make samples suitable for transmission electron microscopy investigations. In comparison, the core NW diameters are about 70 nm.

The interface structure of wz-GaAs/$Pb_{1-x}Sn_xTe$ NWs is investigated using a range of characterization techniques, such as high resolution transmission electron microscopy (HR-TEM), scanning transmission electron microscopy (STEM) and geometric phase analysis (GPA). HR-TEM and STEM images were obtained using FEI Titan 80-300 transmission electron microscope operating at 300 kV with an aberration correction system. The scanning transmission electron microscopy high angle annular dark field detector (STEM−HAADF) images were acquired at the scattering angle range between 80 mrad and 200 mrad camera length, with a converged semi-angle of 9.5 mrad of the incident beam. Geometric phase analysis was performed with Digital Micrograph dislocation density tensor scripts[38-40].

## 3. Results and discussion

Figure 1 shows SEM images of wz-GaAs/$Pb_{1-x}Sn_xTe$ NWs with Sn content x=0.53 and x=0.62. The NW sections with both continuous and fragmented shells are visible. Sample with larger content of Sn in the shell shows more irregular structure due to higher lattice mismatch between the shell and the core materials.

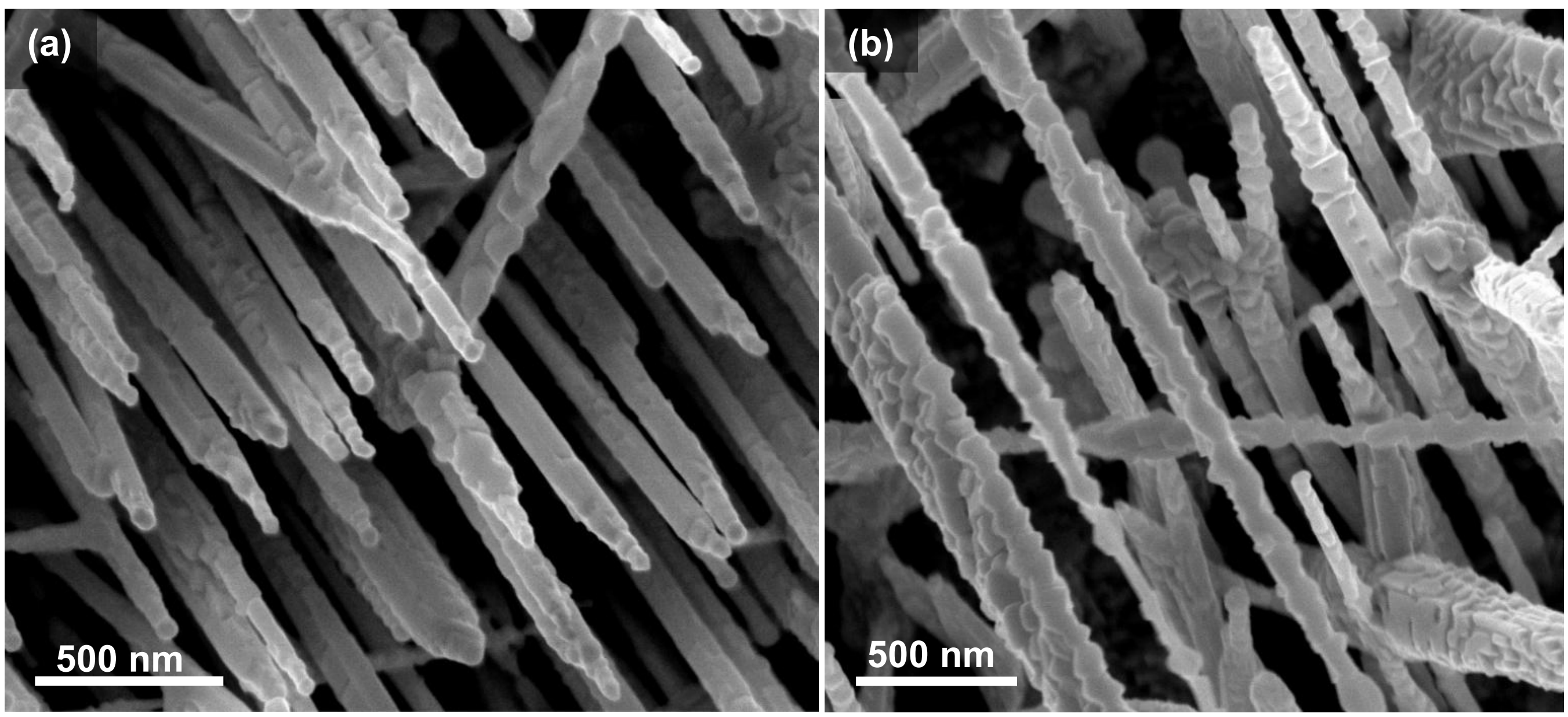

**Fig.1** SEM images of side view of the as-grown wz-GaAs/$Pb_{1-x}Sn_xTe$ x=0.53 (a) and x=0.63 (b) nanowires.

Figure 2 demonstrates close up view on the core-shell interface of wz-GaAs/$Pb_{0.47}Sn_{0.53}Te$ NW. Misfit dislocations at the interface form network with averaged distance between dislocations measured as 15.19 nm. This value can be compared with the predicted value, calculated using a simple equation:

$$d_{dist} = d_{layer} * d_{layer}/(d_{substrate}-d_{layer})$$

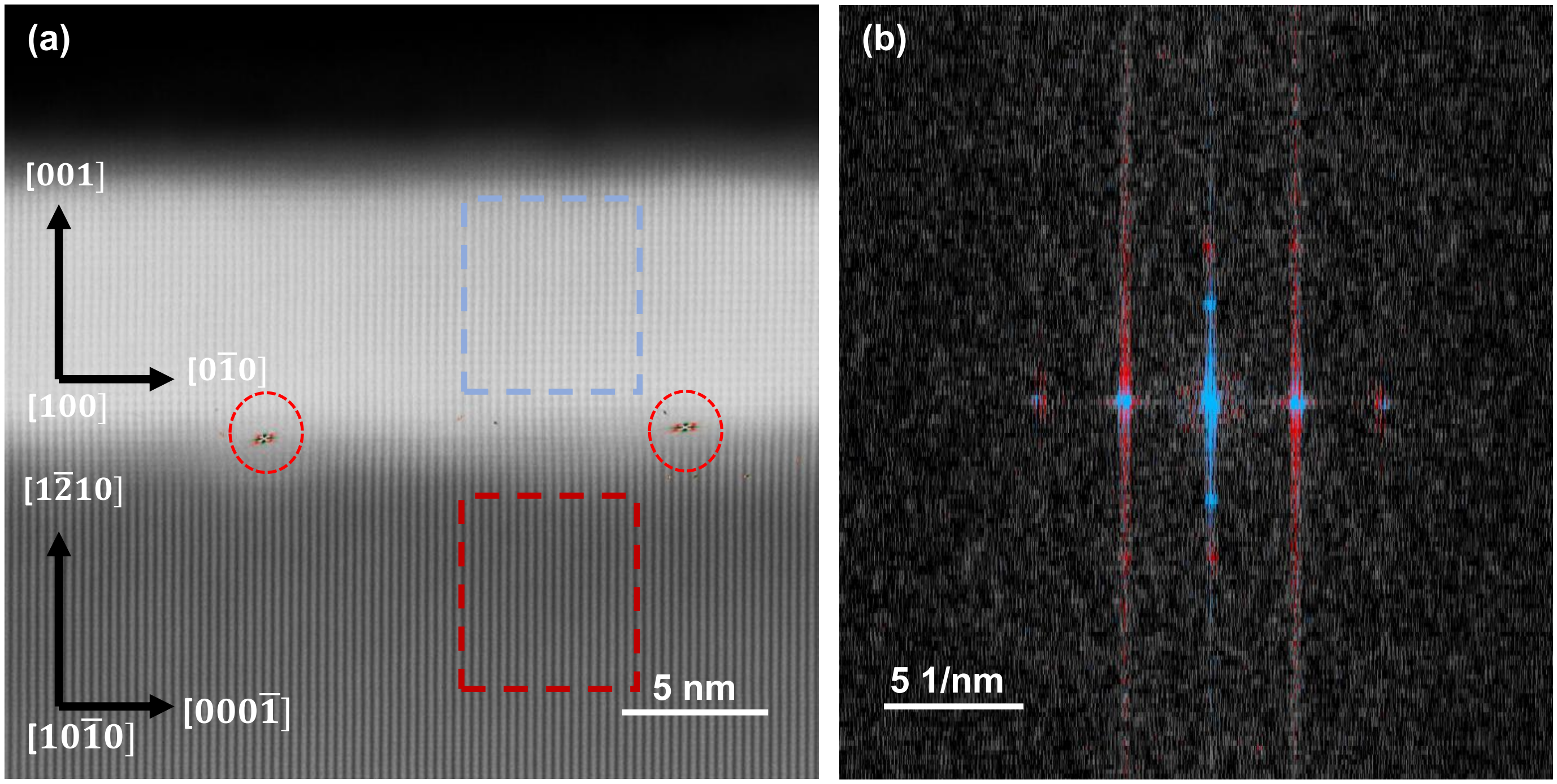

**Fig. 2** (a) - STEM image of a core-shell interface of wz-GaAs/$Pb_{0.47}Sn_{0.53}Te$ nanowire along with superimposed positions of misfit dislocations, marked with red circles; the blue and red dashed squares are the regions selected for FFT images from shell and core respectively (b) FFT image of the shell (blue) and the core (red) sections of the nanowire showing good matching of the lattice parameters of both NW constituents in the axial direction.

where $d_{layer}$ and $d_{substrate}$ are the interplanar distance of the shell and the core materials, respectively, calculated based on HR-TEM images.

Resulting value of 15.1 nm is very similar to the theoretical one suggesting relaxed structure. However, it is worth to notice that dislocation distance varies between different regions of the shell, ranging from around 15 nm near the edge of the shell fragments to around 20 nm deep inside, suggesting that shell structure is less strained near the edge and more strained away from the discontinuous shell fragment. This is illustrated in Figure 3.

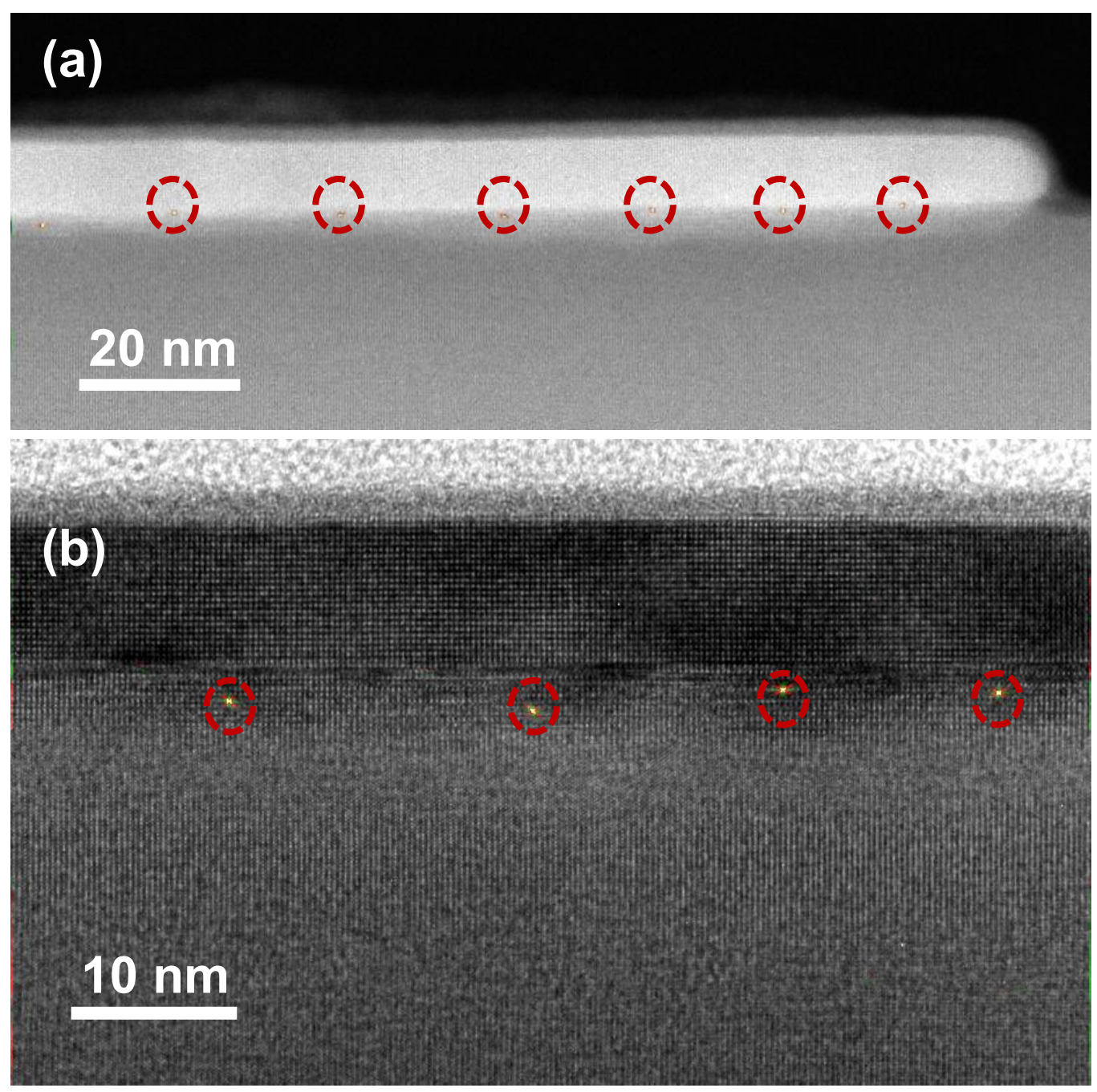

**Fig. 3** (a) STEM and (b) HR-TEM image of wz-GaAs/$Pb_{0.47}Sn_{0.53}Te$ nanowire along with superimposed positions of misfit dislocations, marked with red circles.

Analysis of the acquired TEM and STEM images reveals that the shell consists of several connected fragments rather than a single continuous structure, which contribute to unevenly distributed strain across the shell. This behaviour can be observed also in other samples with discontinuous shells.

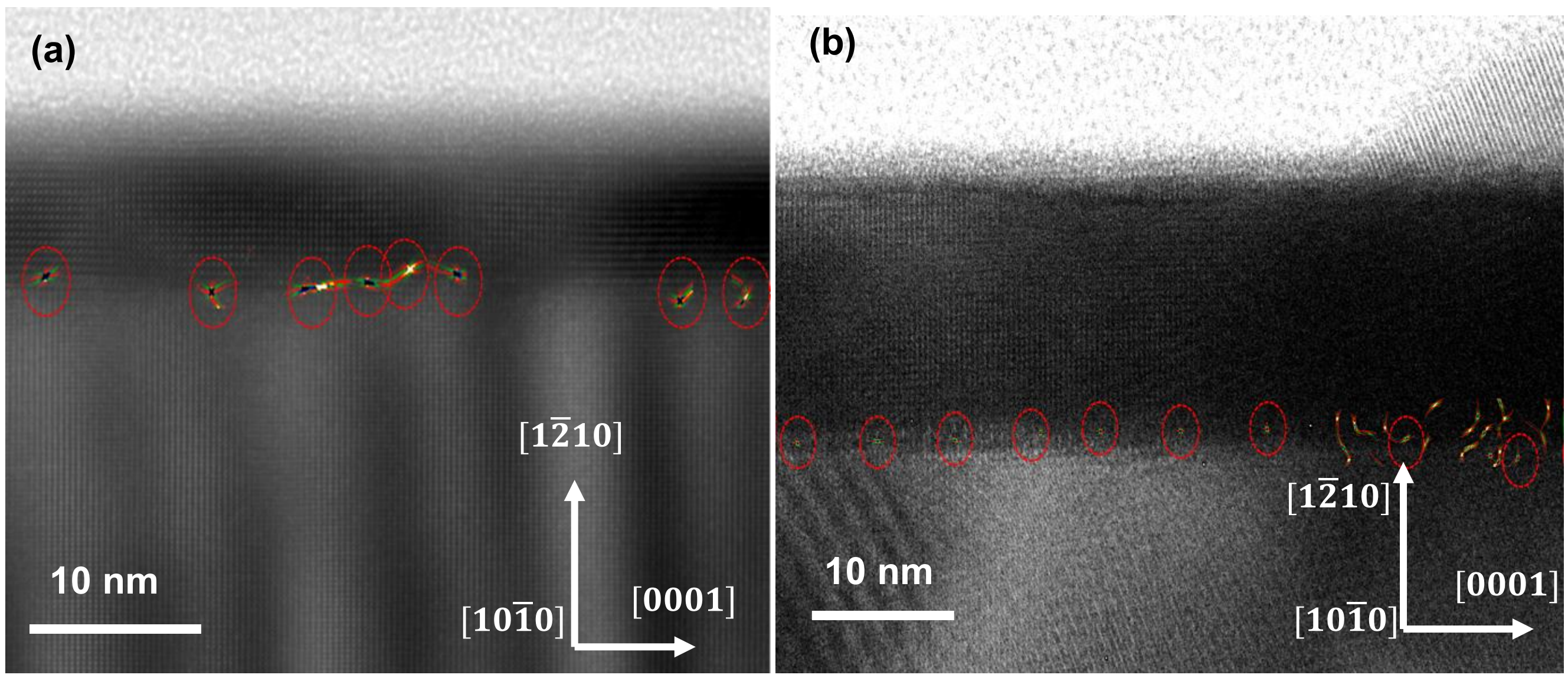

**Fig. 4** HR-TEM images of core/shell interfaces of wz-GaAs/$Pb_{0.37}Sn_{0.63}Te$ (a) and wz-GaAs/PbTe (b) nanowires along with superimposed positions of misfit dislocations, marked with red circles.

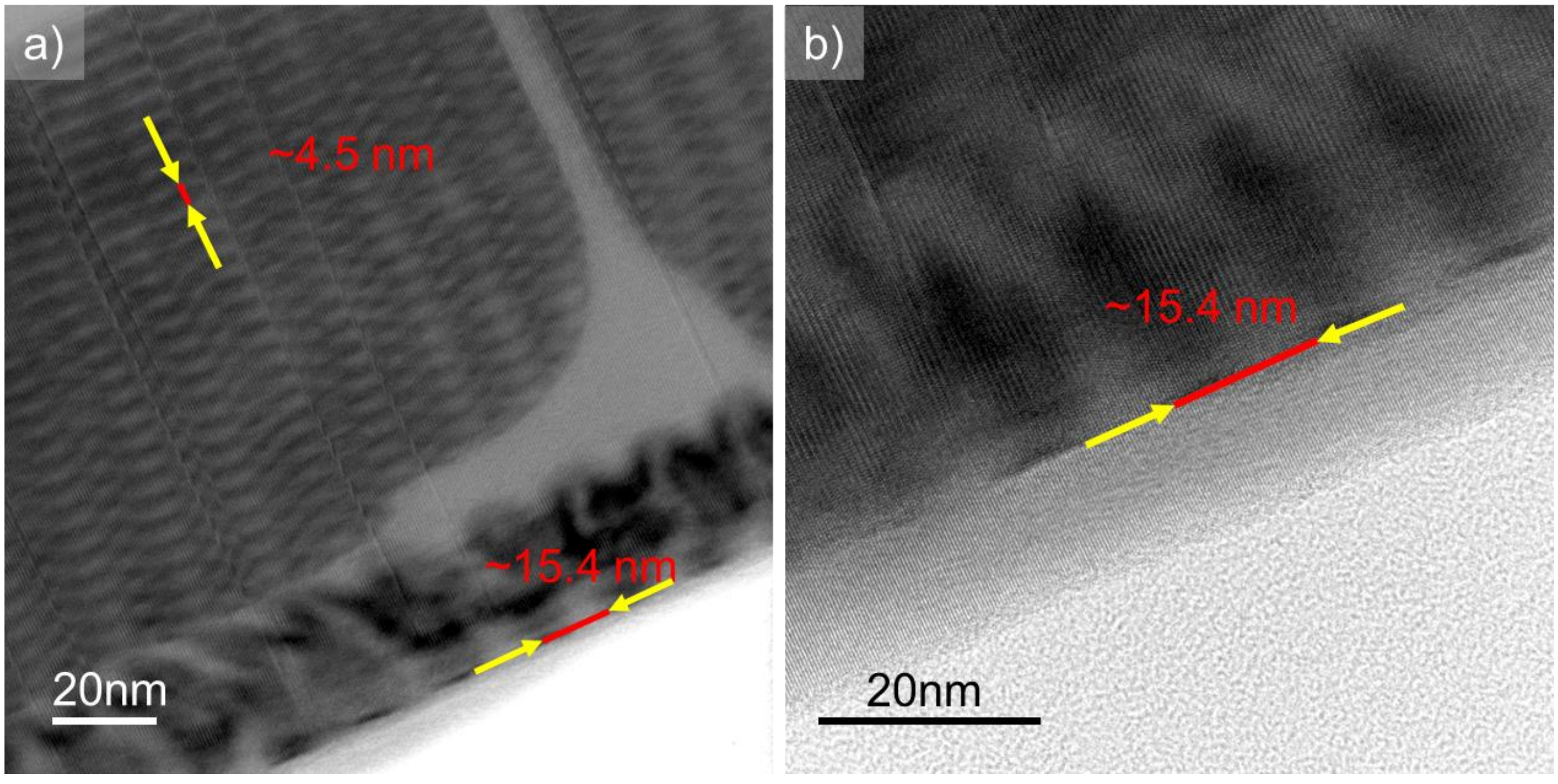

**Fig. 5** HR-TEM images of wz-GaAs/$Pb_{0.47}Sn_{0.53}Te$ core-shell nanowire with moiré patterns shown (a) – perpendicular to (b) - along the core nanowire axis.

HR-TEM images shown in Figure 4 illustrate core/shell interface of wz-GaAs/$Pb_{0.37}Sn_{0.63}Te$ and wz-GaAs/PbTe nanowires. Averaged measured distances between misfit dislocations in the case of wz-GaAs/$Pb_{0.37}Sn_{0.63}Te$ NW are longer than predicted calculated distances, which suggests the presence of residual strain within the structure. Both measured and theoretical dislocation distances for wz-GaAs/$Pb_{0.37}Sn_{0.63}Te$ and wz-GaAs/PbTe are lower than for wz-GaAs/$Pb_{0.47}Sn_{0.53}Te$ NW heterostructure, due to the higher lattice mismatch between the shell and the core. Calculated and measured values for misfit dislocation distances and lattice mismatch are presented in Tab. 1 and Tab. 2. While measuring distance between dislocations in the direction perpendicular to the core was not possible in our experimental setup, a moiré fringes analysis can be performed similar to that for dislocations along the core axis. Values for parallel moiré fringes, distances can be calculated using simple equation, akin to one used for calculating dislocation distances:

$$d_{\text{moire}} = d_{\text{substrate}} * d_{\text{layer}}/(d_{\text{substrate}}-d_{\text{layer}})$$

Calculated moiré fringes distances perpendicular to the core are comparable in investigated samples and amount to between 5.20 nm and 4.75 nm (± 0.1 nm), which is very similar to the observed distances (see Figure. 5 (a)).

Measured averaged distances between moiré fringes in axial direction, as marked in Figure 5, match measured averaged distances between misfit dislocations in the core-shell interfaces rather than the calculated, theoretical values. Thus, moiré patterns analysis provides possibility of initial estimation of strain in the core-shell NW structures, alternative to the geometric phase analysis. Figure 5 (a) also shows moiré fringes in both directions - axial and perpendicular to the core axes. Moiré fringes distances in perpendicular direction are significantly shorter than in the axial one. This is a direct result of much higher mismatch in the lattice parameters of the shell and the core materials in axial and perpendicular directions, as evidenced in FFT patterns shown in Figure 2 (b) and directly visualised in Figure 5.

**Tab. 1** Calculated and measured dislocation and moiré fringes distances and calculated lattice mismatch for presented samples in direction parallel to the core.

| Sample | Measured (Å) | | Calculated (nm) | | Measured dislocation distance (nm) | Simulated moiré distance (nm) | Calculated lattice mismatch | Axial strain (%) |
|---|---|---|---|---|---|---|---|---|
| | $d_{core}$ | $d_{shell}$ | moiré distance | dislocation distance | | | | |
| GaAs/$Pb_{0.47}Sn_{0.53}Te$ | 3.28 | 3.21 | 15.44 | 15.12 | 15.19 | 15.53 | 0.021 | 0.5883 |
| GaAs/$Pb_{0.37}Sn_{0.63}Te$ | 3.28 | 3.19 | 11.39 | 11.07 | 13.78 | 12.23 | 0.028 | 0.2412 |
| GaAs/PbTe | 3.45 | 3.23 | 5.18 | 4.85 | 4.99 | 4.86 | 0.062 | 0.1056 |

**Tab. 2** Calculated dislocation and moiré fringes distances and calculated lattice mismatch for presented samples in direction perpendicular to the core.

| Sample | Measured (Å) | | Calculated (nm) | | Measured moiré distance (nm) | Calculated lattice mismatch | Tangential strain (%) |
|---|---|---|---|---|---|---|---|
| | $d_{core}$ | $d_{shell}$ | moiré distance | dislocation distance | | | |
| GaAs/$Pb_{0.47}Sn_{0.53}Te$ | 1.995 | 3.23 | 5.21 | 8.44 | 4.60 | 0.6208 | 1.2707 |
| GaAs/$Pb_{0.37}Sn_{0.63}Te$ | 1.995 | 3.20 | 5.29 | 8.49 | 4.41 | 0.6062 | 0.7508 |
| GaAs/PbTe | 1.995 | 3.19 | 5.33 | 8.52 | not observed | 0.5977 | 1.3476 |

## 4. Conclusions

We have shown that in spite of the substantial lattice and structural mismatch $Pb_{1-x}Sn_xTe$ can be successfully grown as quasi-continuous shells on the sidewalls of GaAs NWs, which provides opportunity for investigation of topological surfaces of TCIs in the tubular geometry. Measured averaged distance between dislocations for GaAs/$Pb_{1-x}Sn_xTe$ NWs along the high lattice mismatch, tangential [001] direction of the shell are close to the theoretical one, which points to relaxed structure in this direction. In contrast, the corresponding values measured along axial direction are smaller than the theoretical ones, which in turn proves the presence of residual strain within the shell structure. The thorough analysis of the TEM and STEM images reveals that the IV-VI shells consist of several connected fragments rather than a single continuous structure, which contributes to unevenly distributed strain across the shell. Our results show possibility of using moiré patterns analysis as alternative estimation of strain in the core-shell nanowire structures.

### Authors contributions

MW – TEM investigations, strain analysis, paper writing; SD – IV-VI MBE growth, paper writing; PD – IV-VI MBE growth, WZ-P, SK - TEM investigations; WP – III-V MBE growth; JS – III-V MBE growth, conceptualization, financing, paper writing.

**Conflicts of interest**

There are no conflicts to declare.

**Data availability**

The data will be made available upon request.

**Acknowledgements**

We acknowledge the funding from the National Science Centre Poland, project No: 2019/35/B/ST3/03381 and from the “MagTop” project (FENG.02.01-IP.05-0028/23) carried out within the “International Research Agendas” program of the Foundation for Polish Science, co-financed by the European Union under the European Funds for Smart Economy 2021-2027 (FENG). We are grateful for the “Centre of Excellence for nanophotonics, advanced materials and novel crystal growth-based technologies - ENSEMBLE3" project which is carried out within the 2.1 International Research Agendas programme of the Foundation for Polish Science co-financed by the European Union under the European Funds for Smart Economy 2021-2027 (FENG) and the European Union Horizon 2020 research and innovation program Teaming for Excellence (GA. No. 857543) for supporting this work. The publication was created as part of the project of the Minister of Science and Higher Education "Support for the activities of Centers of Excellence established in Poland under the Horizon 2020 program" under contract No. MEiN/2023/DIR/3797.